\documentclass[lettersize,journal]{IEEEtran}

\usepackage{amsmath}
\usepackage{thmtools}
\usepackage{cite}
\usepackage{amssymb,amsfonts}
\usepackage{graphicx}
\usepackage{epstopdf}
\usepackage{textcomp}
\usepackage{multirow}
\usepackage{enumitem}
\usepackage[caption=false,font=normalsize,labelfont=sf,textfont=sf]{subfig}
\usepackage{bbm}
\usepackage{bm}
\usepackage{soul} 
\soulregister{\cite}7
\usepackage{xcolor}
\usepackage{float}
\usepackage{algorithm}
\usepackage{algorithmic}
\newtheorem{theorem}{Theorem}
\newtheorem{lemma}{Lemma}

\usepackage{tikz}
\usepackage{orcidlink,hyperref}
\hypersetup{hidelinks,breaklinks=true}
\usepackage{titlesec}

\begin{document}%

\title{Energy-Aware Random Access Networks: Connection-Based versus Packet-Based}

\author{Anshan Yuan\hspace{0.5mm}\orcidlink{0009-0000-3293-5618}, Fangming Zhao\hspace{0.5mm}\orcidlink{0000-0003-4000-3428}, Xinghua Sun\hspace{0.5mm}\orcidlink{0000-0003-0621-1469},~\IEEEmembership{Member,~IEEE}
\thanks{Anshan Yuan and Xinghua Sun are with the School of Electronics and Communication Engineering, Sun Yat-sen University (Shenzhen Campus), Shenzhen 518107, China (e-mail:\url{yuanansh@mail2.sysu.edu.cn};\url{sunxinghua@mail.sysu.edu.cn})

Fangming Zhao is with the Zhejiang University–University of Illinois Urbana–Champaign Institute,
Zhejiang University, Haining 314400, China (e-mail:\url{fangming.23@intl.zju.edu.cn})}
}

\maketitle
\begin{abstract}
Characterizing and comparing the optimal energy efficiency in energy-aware machine-to-machine (M2M) random access networks remains a challenge due to the distributed nature of the access behavior of nodes.
To address this issue, this letter focuses on the energy efficiency limits of two typical random access schemes, i.e., connection-based Aloha and packet-based Aloha, based on which we conducted a performance comparison. 
Specifically, by integrating limited energy constraints and network throughput, the lifetime throughput can be derived, and further optimized with a guarantee of targeted lifetime via selecting the transmission probability.
Then we present a comparative study on the optimal lifetime throughput of packet-based Aloha and connection-based Aloha to characterize criteria for beneficial connection establishment.
\end{abstract}


\section{Introduction}
\textcolor{black}{The rapid expand of Machine-Type Devices (MTDs) has notably enhanced the prevalence of machine-to-machine (M2M) communications. With the high volume of MTDs, M2M data transmission faces challenges arising from frequent channel access collisions and increased energy consumption. Moreover, in M2M network, nodes are usually battery-operated without recharging ability, resulting in a stringent constraint on energy consumption. Impractical transmission strategies may also shorten the communication lifetime of MTDs, detrimentally impacting the long-term functionality of the network. Consequently, energy efficiency becomes a crucial metric for M2M network\cite{EEsurvey2022, EEsurvey2016}.}

Since the dense MTDs and complicated transmission process in the M2M network result in frequent collisions and unnecessary energy waste, 3GPP recognizes the adoption of MTDs utilizing 2-step (connection-free) and 4-step (connection-based) Random Access Small Data Transmission (RA-SDT) strategies in M2M networks \cite{3GPP_RASDT}. This approach aims to streamline the transmission process to improve throughput and energy efficiency. 
\textcolor{black}{To investigate the throughput performance of the RA-SDT mechanism, the Aloha protocol, known for its simplicity, compared with other existing multiple access technology \cite{noma1,noma2}, still has exhibited significant potential in theoretical exploration and adapting the high volume of MTDs.}
Aloha is divided into two types based on its transmission methodology: packet-based Aloha (PB-Aloha) and connection-based Aloha (CB-Aloha), corresponding to the 2-step RA-SDT and the 4-step RA-SDT, respectively\cite{gao2023random},\cite{Dai_22TON}. 


In PB-Aloha, nodes directly transmit data packets, while in CB-Aloha, nodes initially send a shorter preamble to compete for channel access. 
Intuitively, PB-Aloha incurs higher energy waste and low network throughput in failed data transmissions. Yet, the energy overhead of the preamble and delay created by the connection establishment in CB-Aloha is also nonnegligible. 
The scheme selection largely influences performance in M2M communications. 
{\color{black}Some related works investigated the throughput and delay limits of these two mechanisms. In particular, \cite{19GaoDai} analyzed the optimal throughput of PB-Aloha and CB-Aloha, and showed that the throughput gain brought by connection establishment is significant. \cite{ConnectionAloha} conducted a comparative analysis of the throughput and delay limits between PB-Aloha and CB-Aloha. The beneficial connection establishment threshold in terms of optimal throughput and delay performance was discussed. However, these works ignore the energy constraints when exploring the performance limits, and the optimal configuration may be largely different when considering an energy-aware random access network, which might lead to higher energy consumption and a shorter lifetime of MTDs.}

From the perspective of energy, the energy efficiency of PB-Aloha was optimized using numerical methods in \cite{IRSA_EE}.
Based on the framework in \cite{Aloha},{\color{black}\cite{zhq_EE} derived the energy efficiency limits and conducted a comprehensive analysis of PB-Aloha with the sleeping-awake scheme. The energy efficiency limits of CB-Aloha then largely remain unknown.}
Numerical methods were employed in \cite{CBAloha_TCOMM} to further analyze and compare the energy efficiency of CB-Aloha and PB-Aloha.
However, the numerical comparison in \cite{CBAloha_TCOMM} assumes specific network parameter settings, such as constraining the data payload duration to one time slot.

Since the energy efficiency largely depends on the access protocols and transmission probability, a more reasonable comparison between PB-Aloha and CB-Aloha should be conducted in terms of optimal energy efficiency obtained by tuning transmission probability. This prompts us to characterize the optimal energy efficiency of CB-Aloha first, based on which the performance comparison is further conducted. The main contributions are outlined as follows:

{\color{black}
\begin{itemize}
    \item We accommodate the capability of CB-Aloha to transmit multiple data packets per transmission and propose lifetime throughput to characterize the energy efficiency. Then we derive the closed-form expression of the lifetime throughput, which integrates the node's lifetime, throughput, and energy consumption in each state.
    \item We conduct a comparative lifetime throughput analysis, and present the optimal operating region between PB-Aloha and CB-Aloha, which indicates that CB-Aloha outperforms PB-Aloha in energy efficiency when the network reaches saturation.
    \item Our analysis is also applied to the practical network case, i.e., grant-free 2-step RA-SDT schemes and grant-based 4-step RA-SDT schemes. We present the threshold for packet length that allows 4-step RA-SDT schemes to achieve a better lifetime throughput than 2-step RA-SDT.
\end{itemize}}

\textcolor{black}{The remainder of this letter is organized as follows. 
Section \uppercase\expandafter{\romannumeral2} presents the system model and preliminary analysis. 
Section \uppercase\expandafter{\romannumeral3} characterizes and optimizes the lifetime throughput of CB-Aloha and PB-Aloha. 
The comparative analysis is conducted and applied to the practical RA-SDT case in Section \uppercase\expandafter{\romannumeral4}. 
Finally, Section \uppercase\expandafter{\romannumeral5} concludes the letter.}

\section{System Model and Preliminary Analysis}\label{section:model}
Consider a homogeneous slotted Aloha network\footnote{\textcolor{black}{By using the methodology in \cite{Dai_22TON}, our model can be easily extended to the heterogeneous networks where nodes are divided into several groups according to distinct trafﬁc characteristics and transmission probability.}} where $n$ nodes communicate with a common receiver via a shared channel. Each node has an identical packet arrival rate and possesses an infinite-size buffer\footnote{\textcolor{black}{The performance is not sensitive to the buffer size when it is not too small.}} to accommodate incoming packets. All nodes are synchronized, initiating transmissions only at the beginning of a time slot. Assume the collision model at the receiver, simultaneous transmissions by multiple nodes result in a collision, causing failure for all involved nodes. A transmission is successful only when there are no concurrent transmissions.

\begin{figure}[t]
\vspace{-0.5cm}
\centering
\includegraphics[width=3.49in,height=1.4in]{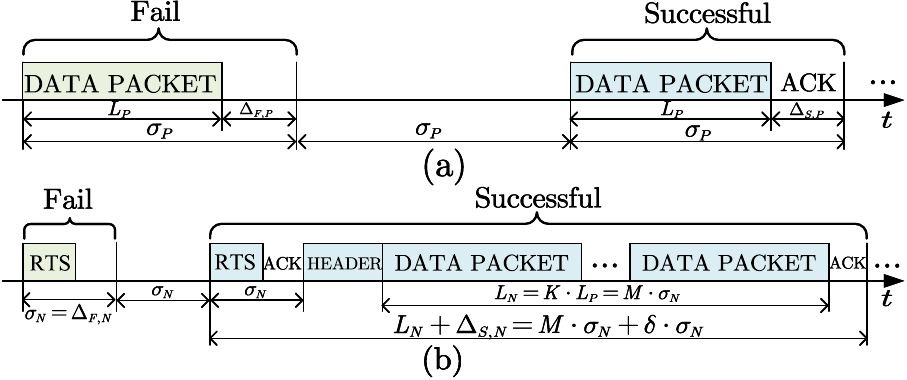}
	\caption{Graphic illustration of (a) PB-Aloha. (b) CB-Aloha.}
	\label{fig5}
	\centering
 \vspace{-0.2cm}
\end{figure}

Fig.\ref{fig5} contrasts the transmission methodologies of PB-Aloha and CB-Aloha, employing superscript and subscript notations for clarity: $N$ denotes CB-Aloha, and $P$ represents PB-Aloha. 
In PB-Aloha, each node transmits a data packet with probability $q$ when its buffer contains a packet, which lasts for one time slot, denoted as $\sigma_P$. When the transmission fails, the node retransmits the data packet with probability $q$ in the next time slot. To standardize the transmission process, the time slot in PB-Aloha is determined by the data packet length. 

Conversely, in CB-Aloha, each node sends a short length request (RTS) for channel competition, that lasts for $\sigma_N$ with probability $q$ to establish a connection once its buffer accumulates $K$ data packets. Note that the length of RTS is smaller than the data packet, i.e., $\sigma_N < \sigma_P$, then the unit time slot length in CB-Aloha is determined by the length of RTS. A successful RTS reception prompts an acknowledgment (ACK) from the receiver, initiating the data packet transmission, and each transmission comprises $K$ data packets transmission spanning $M$ time slots and the signaling overhead lasting $\delta$ time slots. Failed attempts lead to retransmission of the RTS in the subsequent time slot with probability $q$. 

The CB-Aloha can be analyzed by using a request-queue model\cite{ConnectionAloha}. 
As each node generates a request and competes for the channel with this request, the network performance hinges on the aggregate behavior of the Head-Of-Line (HOL) request. 
\textcolor{black}{By establishing the state transition process of each HOL request, the probability of successful HOL request transmission $p_N$ in both saturated (i.e., each CB-Aloha node always has a packet to transmit) and unsaturated conditions has been characterized in \cite{ConnectionAloha} as:}

\vspace{-5pt}
\begin{footnotesize}
\begin{equation}\label{PQ_Cor}
p_N \!=\!\!\left\{\!\!\!\!
\begin{array}{ll}
\exp\left(  {\it {\mathbb W}}_{\text{0}}(-\frac{\hat{\lambda}_N}{M-\hat{\lambda}_N(M+\delta-1)}) \right)  \,\,\,\,\text{if} \,\, q\in \\
\,\,\,\,\,\,\,\left[ -\frac{1}{n}{\it {\mathbb W}}_{0}(-\frac{\hat{\lambda}_N}{M-\hat{\lambda}_N(M+\delta-1)}), \right. \left. -\frac{1}{n}{\it {\mathbb W}}_{-1}(-\frac{\hat{\lambda}_N}{M-\hat{\lambda}_N(M+\delta-1)})\right]\\
\exp\left( -nq\right) \,\,\,\,\,\,\,\,\,\,\,\,\,\,\,\,\,\,\,\,\,\,\,\,\,\,\,\,\,\,\,\,\,\,\,\,\,\,\,\,\,\,\,\,\,\,\,\,\,\,\,\,\,\,\text{otherwise,}\\
\end{array}\right.
\end{equation}
\end{footnotesize}

\noindent
where ${\it {\mathbb W}}_{0}(\cdot)$ and ${\it {\mathbb W}}_{-1}(\cdot)$ are two branches of the Lambert $\mathbb{W}$ function. The node throughput is given by \cite{ConnectionAloha}:
\begin{equation}\label{th_gen}
\lambda_{\text{out}}^N=\left\{\!\!\!
\begin{array}{ll}
\lambda_N \,\,\,\,\,\,\,\,\,\,\,\,\,\,\,\,\,\,\,\,\,\,\,\,\,\,\,\,\,\,\,\,\,\,\,\,\,\,\,\,\,\,\,\,\,\,\,\,\,\,\,\text{if}\,\,  p_N\in S(p_N,\hat{\lambda}_N) \\
\frac{M}{\frac{n}{-p_N \ln{p_N}}+n(M+\delta-1)}
\,\,\,\,\,\,\,\,\text{otherwise,}\,\,\,\,\,\,
\end{array}\right.
\end{equation}
where
\begin{small}
    \begin{equation}
	\begin{array}{ll}
		S(p_N,\hat{\lambda}_N)=\left\lbrace p_N:\exp{\left( {\it {\mathbb W}}_{-1}(-\frac{\hat{\lambda}_N}{M-\hat{\lambda}_N(M+\delta-1)})\right) } < \right. \\
		 \left. ~~~~~~~~~~~~~~~~~~~~~~~p_N< \exp{\left( {\it {\mathbb W}}_{0}(-\frac{\hat{\lambda}_N}{M-\hat{\lambda}_N(M+\delta-1)})\right) }\right\rbrace .
	\end{array}
\end{equation}
\end{small}

Assume that each node possesses an initial finite amount of energy $E$, consequently determining its finite lifetime.
For homogeneous CB-Aloha, the expected lifetime $T_N$ is identical for each node, in the unit of time slots. Throughout the lifetime, each node can exist in four states: (1) \textit{failed state}, i.e., the node attempts to transmit a request but fails. (2) \textit{successful state}, i.e., the node successfully transmits both request and data. (3) \textit{waiting state}, i.e., the node holds one virtual HOL request, awaiting channel access. (4) \textit{idle state}, i.e., the node's request queue is devoid of any pending requests.

Let $n_{W}$, $n_{I}$, $n_{F}$ and $n_{S}$ denote the expected number of requests for each node being in the waiting, idle, failed and successful state during its lifetime, respectively. 
The successful transmission state lasts for $M+\delta$ time slots while the other states just last for 1 time slot. The expected lifetime $T_N$, in the unit of time slots $\sigma_N$, can be written as:
\begin{equation}\label{Eq2}
T_N=n_{I}+n_{W}+n_{F}+n_{S}(M+\delta).
\end{equation}
{\color{black} Let $P_{I}$, $P_{W}$, and $P_{T}$ denote the normalized power consumption in the idle, waiting, and transmission states, respectively, as determined by the practical network configuration.
Generally, we assume that $P_{I}=P_{W}\leq P_{T}$.} According to the total energy constraint of each node, we have
\begin{align}\label{Eq3}
P_{W}(n_{I}+ n_{W})+P_{T}\left( n_{F}+n_{S}\left( M+\delta\right) \right) =E/\sigma_N.
\end{align}

Due to the constraint of finite energy, each node has a restricted capacity to deliver packets throughout its lifetime.
The lifetime throughput of each node $U_N$ is defined as \textit{the average number of packets that a node can successfully transmit during its entire lifetime.}
In the following section, $U_N$ will be characterized and further optimized.

\section{Lifetime Throughput limits Analysis}
\subsection{Lifetime Throughput limits in CB-Aloha}
This subsection aims to obtain the maximum lifetime throughput of each CB-Aloha node $U_{\max}^{p,N}$. 
The lifetime throughput of each node $U_N$, can be written as:
\begin{equation}\label{Eq4}
	\setlength\belowdisplayskip{3pt}
	\setlength\abovedisplayskip{3pt}
U_N=\lambda_{\text{out}}^N T_N.
\end{equation}

The following lemma presents the expression of the expected lifetime of each CB-Aloha node $T_N$.

\begin{lemma}\label{lemma1}
The expected lifetime of each node is given by
\begin{equation}\label{T_gen}
T_N=\left\{\!\!\!
\begin{array}{ll}
\frac{E/\sigma_N}{\frac{\left( 1+(M+\delta-1)p_L\right) \lambda_N}{Mp_L}(P_{T}-P_{W})+P_{W}}
\,\,\,\,\,\,\,\text{if}\,\,  p_N\in S(p_N,\hat{\lambda}_N)\\
\frac{\frac{E}{\sigma_N}(1-p_N\ln{(p_N)}(M+\delta-1))}{P_{W}-\frac{((n-1)P_{W}+P_{T})(M+\delta-1)p_N\ln{(p_N)}}{n}-\frac{(P_{T}-P_{W})\ln{p_N}}{n}}\\
~~~~~~~~~~~~~~~~~~~~~~~~~~~~~~~~~~~~~~~\text{otherwise,}\,\,\,\,\,\,\\
\end{array}\right.
\end{equation}
where $p_L=\exp{\left({\it {\mathbb W}}_{0}\left(-\frac{\hat{\lambda}_N}{M-\hat{\lambda}_N(M+\delta-1)}\right)\right) }.$
\end{lemma}

\begin{IEEEproof}
Please see Appendix A.
\end{IEEEproof}
By combining \eqref{th_gen}, \eqref{Eq4} and \eqref{T_gen}, the lifetime throughput of each node $U_N$ is given by
\begin{equation}\label{M_gen}
U_N \!=\!\!\left\{\!\!\!\!\!
\begin{array}{ll}
\frac{E/\sigma_N}{\frac{ 1+(M+\delta-1)p_L }{Mp_L}(P_{T}-P_{W})+\frac{P_{W}}{\lambda_N}}\,\,\,\,\,\,\,\,\,\,\,\,\,\,\,\text{if}\,\, p_N\in S(p_N,\hat{\lambda}_N) \\
\frac{E/\sigma_N}{\frac{((n-1)P_{W}+P_{T})(M+\delta-1)}{M}+\frac{P_{T}-P_{W}}{Mp_N}-\frac{nP_{W}}{Mp_N\ln{p_N}}}
\,\,\,\,\,\text{otherwise.}\,\,\,\,\,\,
\end{array}\right.
\end{equation}

\begin{figure}[t]
	\centering
	\vspace{-0.7cm}
	\includegraphics[width=2.8in,height=1.7in]{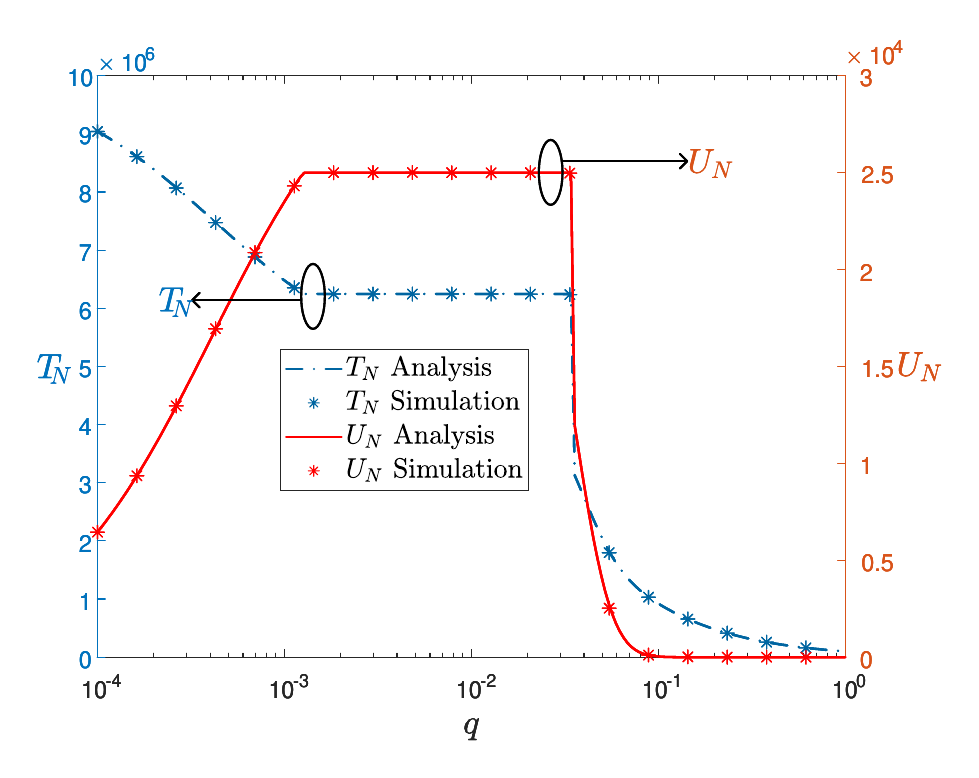}
	\caption{Lifetime $T_N$ and lifetime throughput $U_N$ of each node, $n=100$, $M=8$, $\delta=4$, $E/\sigma_N=10^7$, $P_T=100$, $P_W=1$, ${\lambda}_N=0.004$.}
	\label{simulation}
	\centering
	\vspace{-0.7cm}
\end{figure}
Fig. \ref{simulation} shows the simulation results about lifetime $T_N$ and lifetime throughput $U_N$ versus transmission probability $q$ of each node. 
Our simulation considers a $n=100$ nodes CB-Aloha network where each node has Bernoulli packet arrival rate $\lambda_N=0.004$. Given the power consumption $P_T$ and $P_W$ and a finite amount of energy normalized by time slot $E/\sigma_N$, we calculate the lifetime $T_N$ and the average number of packets that a node can successfully transmit during its lifetime as their lifetime throughput $U_N$.
\textcolor{black}{Both simulations and the analytical result of Eq. \eqref{T_gen} and Eq. \eqref{M_gen} have shown that as $q$ increases, $T_N$ is non-increasing while $U_N$ is non-monotonic, indicating a trade-off between $T_N$ and $U_N$ when $q$ is small.}
We are interested in maximizing the lifetime throughput of each node $U_N$ by tuning the transmission probability $q$. In practice, each node is expected to live longer than a certain threshold value to avoid early death. \textcolor{black}{Under such lifetime constraint, we have the following constrained optimization problem:}
\textcolor{black}{
\begin{align}\label{Mc_did1}
U_{\max}^{p,N}=&\max_{q} \;\; {U_N}\\
s.t. \;\;\;&T_N\ge T_0^N.\notag
\end{align}}
The following theorem gives the solution to problem \eqref{Mc_did1}.

\begin{theorem}\label{t1}
Given the data length $M$, the maximum lifetime throughput $U_{\max}^{p,N}=\max\limits_{q} U_N$ under the constraint of $T_N\ge T_0^N$ is given by
\begin{equation}\label{connection_Umaxp}
U_{\max}^{p,N}=\left\{\!\!\!
\begin{array}{ll}
\frac{E/\sigma_N}{\frac{ 1+(M+\delta-1)p_{L} }{Mp_{L}}(P_{T}-P_{W})+\frac{P_{W}}{\lambda_N}}\\
\,\,\,\,\,\,\,\,\,\,\,\,\,\,\,\,\,\,\text{if}\,\,\,\,\lambda_N\leq\lambda_{M}^N\,\,\text{and}\,\, T_0^N\le T_{0}^{*,N}\\
\frac{E/\sigma_N}{\frac{((n-1)P_{W}+P_{T})(M+\delta-1)}{M}+\frac{P_{T}-P_{W}}{Mp_\text{m}}-\frac{nP_{W}}{Mp_\text{m}\ln{p_\text{m}}}}\\
\,\,\,\,\,\,\,\,\,\,\,\,\,\,\,\,\,\,\text{if}\,\,\,\,\lambda_N>\lambda_{M}^N\,\,\text{and}\,\, T_0^N\le T_{0}^{*,N}\\
\frac{E/\sigma_N}{\frac{((n-1)P_{W}+P_{T})(M+\delta-1)}{M}+\frac{P_{T}-P_{W}}{Mp_{c}^N}-\frac{nP_{W}}{Mp_{c}^N\ln{p_{c}^N}}}\\
\,\,\,\,\,\,\,\,\,\,\,\,\,\,\,\,\,\,\text{if}\,\,\,\,T_{0}^{*,N}< T_{0}^N \leq \frac{E/\sigma_N}{P_{W}} ,\\
\end{array}\right.
\end{equation}
otherwise, \eqref{Mc_did1} has no feasible solution. \textcolor{black}{The corresponding optimal transmission probability $q_{\max}^N$ is set to be}
\begin{small}
\begin{equation}\label{qM_N}
\textcolor{black}{
q_{\max}^N=\left\{\!\!
\begin{array}{ll}
\left[ \frac{-{\it {\mathbb W}}_{0}(-\frac{\hat{\lambda}_N}{M-\hat{\lambda}_N(M+\delta-1)})}{n},  \frac{-{\it {\mathbb W}}_{-1}(-\frac{\hat{\lambda}_N}{M-\hat{\lambda}_N(M+\delta-1)})}{n}\right] \\ \,\,\,\,\,\,\,\,\,\,\,\,\,\,\,\,\,\,\,\,\,\,\,\,\,\,\,\,\,\,\,\,\,\,\,\,\,\,\,\,\,\,\text{if}\,\,\,\,\lambda_N\leq\lambda_{M}^N\,\,\text{and}\,\, T_0^N\le T_{0}^{*,N}\\
-(\ln{p_\text{m}})/n\,\,\,\,\,\,\,\,\,\,\,\,\,\text{if}\,\,\,\,\lambda_N>\lambda_{M}^N\,\,\text{and}\,\, T_0^N\le T_{0}^{*,N}\\
-(\ln{p_c^N})/n \,\,\,\,\,\,\,\,\,\,\,\,\text{if}\,\,\,\,T_{0}^{*,N} < T_{0}^N \leq \frac{E/\sigma_N}{P_{W}}.\\
\end{array}\right.
}
\end{equation}
\end{small}

\noindent
where $\lambda_M^N$ marks the boundary of the saturated region ($\lambda_N>\lambda_{M}^N$) and the unsaturated region ($\lambda_N\leq\lambda_{M}^N$) in CB-Aloha, which is given by
\begin{equation}
 \lambda_M^N =\frac{M}{n}
\frac{p_\text{m}\ln{p_\text{m}}}{p_\text{m}\ln{p_\text{m}}\left( M+\delta-1 \right)-1 } ,
\end{equation}
$T_{0}^{*,N}$ can be expressed as
\begin{equation}
 T_{0}^{*,N}=\max\left\{ T_N\left( p_{L} \right),T_N\left( p_\text{m} \right) \right\},  
\end{equation}
$p_\text{m}$ is given by 
\begin{equation}\label{eq:pm}
  p_\text{m}=\exp\left(\tfrac{n-\sqrt{n^2+4n\left( \frac{P_{T}}{P_{W}}-1 \right)}}{2\left( \frac{P_{T}}{P_{W}}-1 \right) } \right),
\end{equation}
and $p_c^N$ is the solution of the following equation:
\begin{equation}\label{solvePc}
	\begin{array}{l}	T_{0}^N=\frac{\frac{E}{\sigma_N}(1-p_{c}^N\ln{p_c^N}(M+\delta-1))}{P_{W}-\frac{((n-1)P_{W}+P_{T})(M+\delta-1)p_{c}^N\ln{p_{c}^N}}{n}-\frac{(P_{T}-P_{W})\ln{p_c^N}}{n}}.
	\end{array}
\end{equation}

\end{theorem}

\begin{IEEEproof}
Please see Appendix B.
\end{IEEEproof}

\begin{figure}[t]
	\centering
	\vspace{-0.6cm}
	\includegraphics[width=3.49in,height=1.8in]{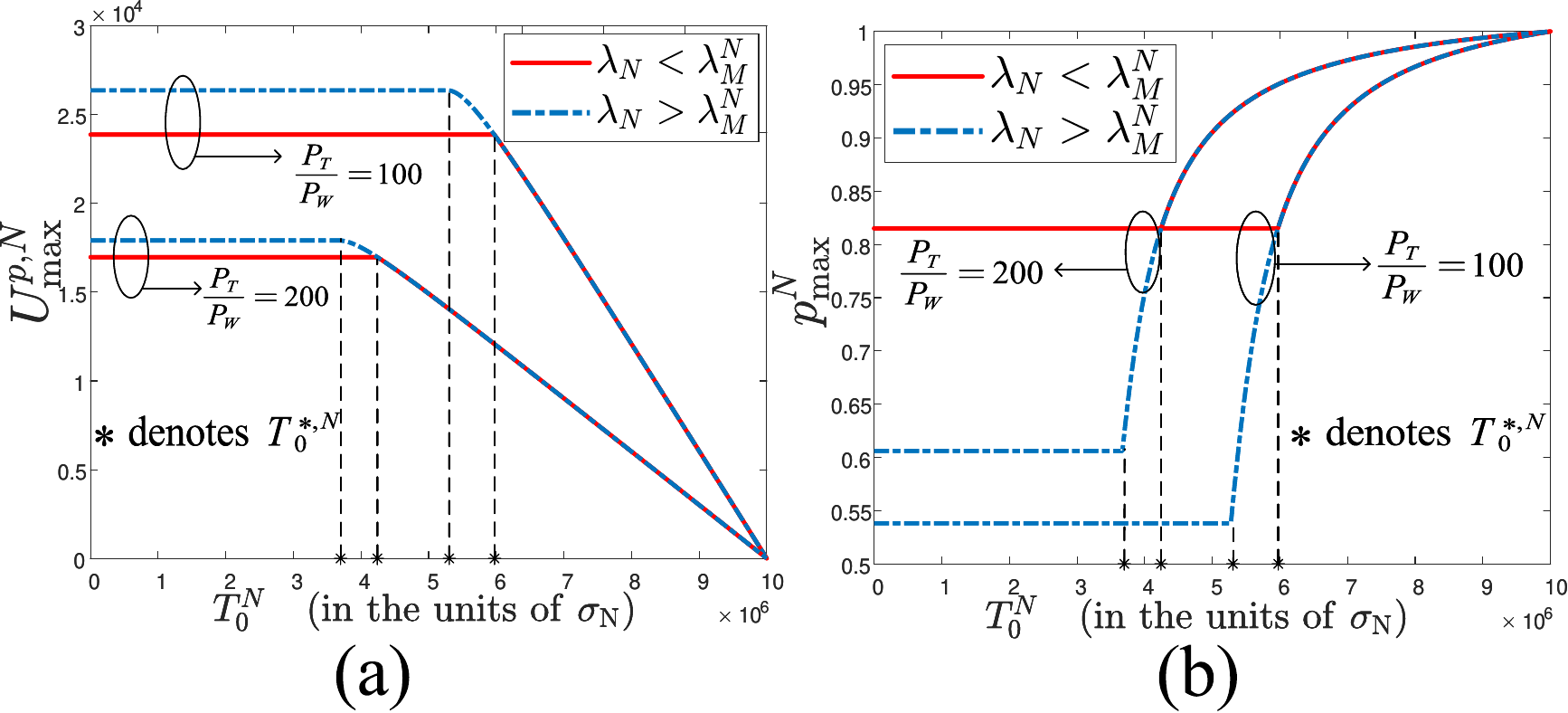}
	\caption{The (a) maximum lifetime throughput $U_{\max}^{p,N}$ and (b) optimal successful transmission probability $p_{\max}^N$ versus constraint $T_0^N$ in different packet arrival rate $\lambda_N$, $n=100$, $M=6$, $\delta=4$, $E/\sigma_N=10^7$, $P_W=1$.}
	\label{showTheorem_1}
	\centering
	\vspace{-0.5cm}
\end{figure}



\textcolor{black}{Fig. \ref{showTheorem_1}a depicts the maximum lifetime throughput $U_{\max}^{p,N}$ versus lifetime constraint $T_0^N$.
It illustrates that when $T_0^N<T_0^{*,N}$, $U_{\max}^{p,N}$ is constant for $T_0^N$ which aligns the case that without lifetime constraint. As $T_0^N>T_0^{*,N}$, $U_{\max}^{p,N}$ decreases when $T_0^N$ increases and $U_{\max}^{p,N}$ and is unaffected by the arrival rate $\lambda_N$ since the network becomes saturated. Additionally, Fig. \ref{showTheorem_1}b indicates that the optimal transmission probability $p_{\max}^N$ remains unchanged for $T_0^N<T_0^{*,N}$. Above $T_0^{*,N}$, or with higher transmission power consumption $P_T$, nodes lower their transmission probability $q$ to meet the constraints, which results in an increased $p_{\max}^N$ with higher $T_0^N$ or $P_T$.}

\textcolor{black}{\subsection{Lifetime Throughput Limits in PB-Aloha}}
The lifetime throughput of PB-Aloha can be directly obtained by that of CB-Aloha.
Specifically, in PB-Aloha, the data packet reduces to unit-time slot for each transmission, and there is no additional overhead required for establishing the connection. By substituting $M=1$ and $\delta=0$ into \eqref{T_gen} and \eqref{M_gen}, the expected lifetime $T_P$ and lifetime throughput $U_P$ of each PB-Aloha node can be obtained as following:
\begin{equation}
	T_P=\left\{\!\!\!
	\begin{array}{ll}
	\frac{E/\sigma_P}{\frac{\lambda_P}{\exp\left( \mathbb{W}_0(-n\lambda_P) \right) }(P_T-P_W)+P_W}\,\,\text{if}\,\,\,  p\in S(p,\hat{\lambda}_P)\\
		\frac{E/\sigma_P}{P_{W}-\frac{(P_{T}-P_{W})\ln{p}}{n}}	\,\,\,\,\,\,\,\,\,\,\,\,\,\,\,\,\,\,\,\,\,\,\,\,\,\,\,\text{otherwise,}\,\,\,\,\,\,\\
	\end{array}\right.
\end{equation}
\begin{equation}
	U_P=\left\{\!\!\!
	\begin{array}{ll}
		\frac{E/\sigma_P}{\frac{ (P_{T}-P_{W}) }{\exp{\left( {\it {\mathbb W}}_{0}(-n\lambda_P)\right) }}+\frac{P_{W}}{\lambda_P}}
		\,\,\,\,\,\,\,\,\,\,\,\,\,\,\,\text{if}\,\,\, p\in S(p,\hat{\lambda}_P)\\
		\frac{E/\sigma_P}{\frac{P_{T}-P_{W}}{p}-\frac{nP_{W}}{p\ln{p}}}
		\,\,\,\,\,\,\,\,\,\,\,\,\,\,\,\,\,\,\,\,\,\,\,\,\,\,\,\text{otherwise,}\,\,\,\,\,\,
	\end{array}\right.
\end{equation}
where 
\begin{small}
    \begin{equation}
        S(p,\hat{\lambda}_P)
            =\left\{ p:\exp{\left( {\it {\mathbb W}}_{-1}(-n\lambda_P)\right)}< p < \exp{\left( {\it {\mathbb W}}_{0}(-n\lambda_P)\right) }\right\}.   
    \end{equation}
\end{small}

\noindent
Therefore, the maximum lifetime throughput $U_{\max}^{p,P}=\max\limits_{q} U_P$ under the constraint of $T_P\ge T_0^P$ is given by
\begin{equation}\label{packet_Umaxp}
	U_{\max}^{p,P} \!\!=\left\{\!\!\!\!
	\begin{array}{ll}
		\frac{E/\sigma_P}{\frac{ (P_{T}-P_{W}) }{\exp{\left( {\it {\mathbb W}}_{0}(-n\lambda_P)\right) }}+\frac{P_{W}}{\lambda_P}}     \,\,\,\,\,\text{if}\,\,\,\,\lambda_P\leq\lambda_{M}^P\,\,\text{and}\,\, T_0^P\le T_{0}^{*,P}\\
		\frac{E/\sigma_P}{\frac{ (P_{T}-P_{W}) }{\exp{\left( {\it {\mathbb W}}_{0}(-n\lambda_{M}^P)\right) }}+\frac{P_{W}}{\lambda_{M}^P}}\,\,\,\,\text{if}\,\,\,\,\lambda_P>\lambda_{M}^P\,\,\text{and}\,\, T_0^P\le T_{0}^{*,P}\\
		\frac{E/\sigma_P}{\frac{P_{T}-P_{W}}{p_{c}^P}-\frac{nP_{W}}{p_{c}^P\ln{p_{c}^P}}}\,\,\,\,\,\,\,\,\,\,\,\text{if}\,\,\,\,T_{0}^{*,P} < T_{0}^P \leq \frac{E/\sigma_P}{P_{W}}.\\
	\end{array}\right.
\end{equation}
\textcolor{black}{The corresponding optimal transmission probability $q_{\max}^{P}$ is given by
\begin{equation}\label{qM_P}
q_{\max}^P=\left\{\!\!\!\!
\begin{array}{ll}
\left[ -\frac{1}{n}{ {\mathbb W}}_{0}(-n\lambda_P), -\frac{1}{n}{\it {\mathbb W}}_{-1}(-n\lambda_P)\right] \\ \,\,\,\,\,\,\,\,\,\,\,\,\,\,\,\,\,\,\,\,\,\,\,\,\,\,\,\,\,\,\,\,\,\text{if}\,\,\,\,\lambda_P\leq\lambda_{M}^P\,\,\text{and}\,\, T_0^P\le T_{0}^{*,P}\\
-(\ln{p_\text{m}})/n\,\,\,\,\text{if}\,\,\,\,\lambda_P>\lambda_{M}^P\,\,\text{and}\,\, T_0^P\le T_{0}^{*,P}\\
-(\ln{p_c^P})/n \,\,\,\,\text{if}\,\,\,\,T_{0}^{*,P} < T_{0}^P \leq \frac{E/\sigma_P}{P_{W}}.\\
\end{array}\right.
\end{equation}}

\noindent
where $\lambda_M^P$ marks the boundary of the saturated region ($\lambda_P>\lambda_{M}^P$) and the unsaturated region ($\lambda_P\leq\lambda_{M}^P$) in PB-Aloha, and can be expressed as
\begin{equation}
\lambda_{M}^P= \tfrac{\sqrt{1+\frac{4}{n}\left( \frac{P_{T}}{P_{W}}-1 \right)}-1}{2\left( \frac{P_{T}}{P_{W}}-1 \right) } \exp\left\{ \tfrac{n-\sqrt{n^2+4n\left( \frac{P_{T}}{P_{W}}-1 \right)}}{2\left( \frac{P_{T}}{P_{W}}-1 \right) } \right\}. 
\end{equation}
$T_{0}^{*,P}$ can be expressed as
\begin{equation}
 T_{0}^{*,P}=\frac{E/\sigma_P}{\frac{\min\left\lbrace \lambda_P,\lambda_{M}^P\right\rbrace }{\exp\left\lbrace \mathbb{W}_0(-n\min\left\lbrace \lambda_P,\lambda_{M}^P \right\rbrace ) \right\rbrace } (P_T-P_W)+P_W},
\end{equation}
$p_{\textrm{m}}$ has been expressed in \eqref{eq:pm}, and $p_{c}^P$ is given by
\begin{equation}
    p_{c}^P=\exp\left\lbrace -n\frac{\frac{E}{\sigma_P T_0^P}-P_W}{P_T-P_W} \right\rbrace.
\end{equation}

\section{Energy-Aware M2M: CB-Aloha or PB-Aloha}
In this section, we will compare the optimal lifetime throughput of the CB-Aloha and PB-Aloha protocols to ascertain the optimal operating regime of these two schemes in energy-aware M2M communication.

\begin{figure}[t]
	\vspace{-1cm}
	\centering
 	\subfloat[]{\includegraphics[width=1.70in,height=1.6in]{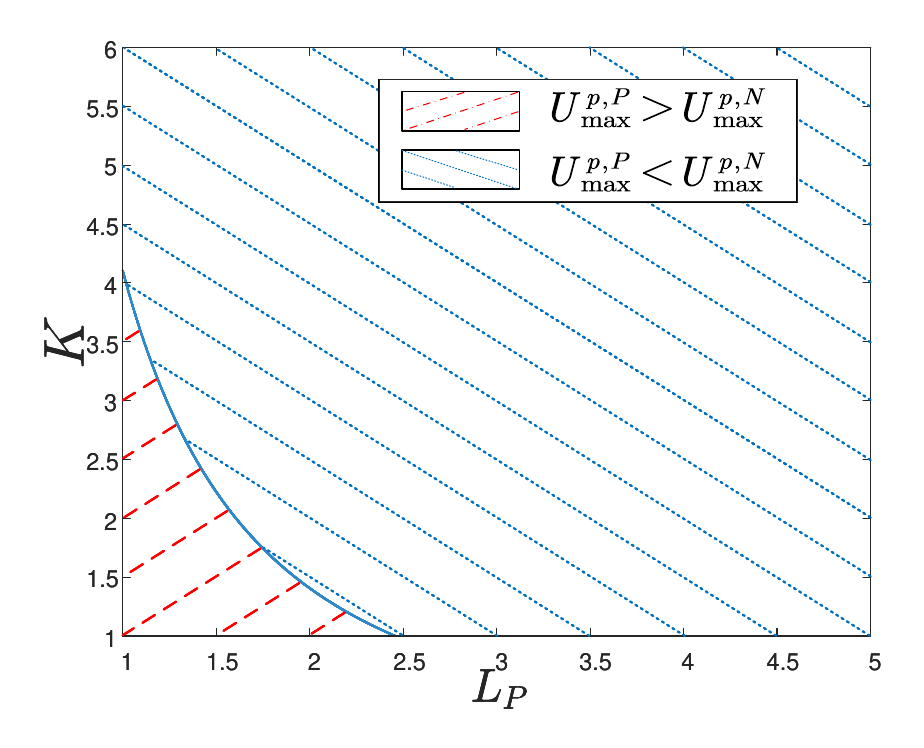}
		\label{Both_network_saturated}}
	\subfloat[]{\includegraphics[width=1.70in,height=1.6in]{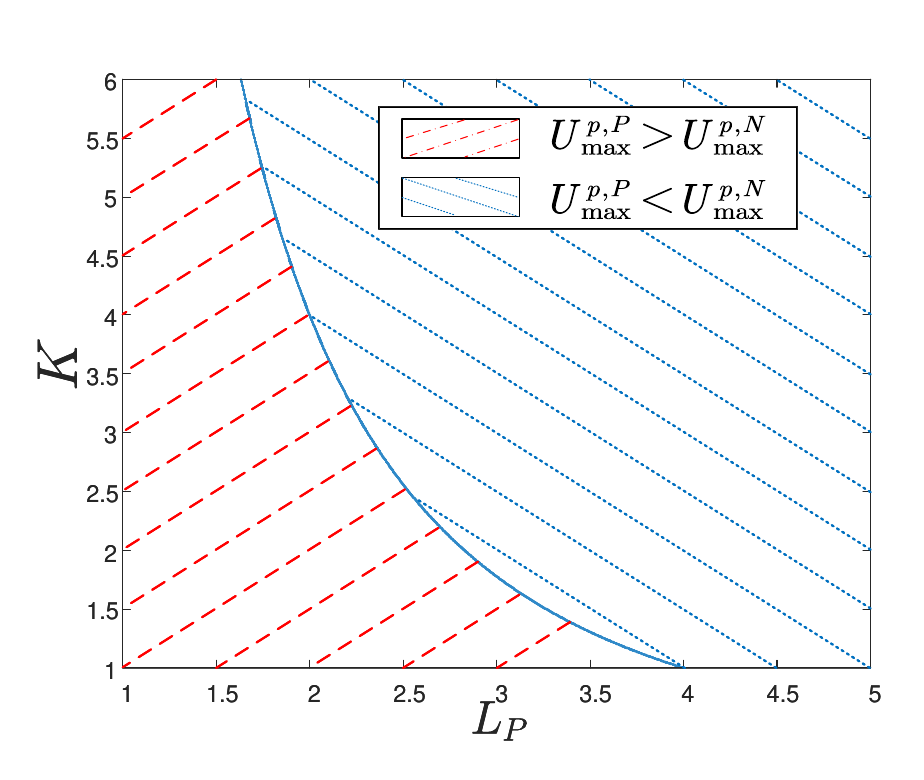}
		\label{Both_network_unsaturated}}
	\caption{Comparison about $U_{\max}^{p,P}$ and $U_{\max}^{p,N}$ in different $K,L_P$. $n=100$, $\delta=4$,  $P_{T}=100$, $P_{W}=1$, $\Delta_{S,P}=2$, $T_0^N=T_0^P=0$, $\sigma_N=2$, \textcolor{black}{(a) Both networks are saturated . (b) Both networks are unsaturated.}}
	\vspace{-0.5cm}
	\label{Both_network_compare}
\end{figure}

\subsection{Performance Comparison}

As Fig. \ref{fig5}b illustrates, let $L$, $\Delta_S$, and $\Delta_F$ denote the length of the data payload and the overhead time for each successful or failed transmission.
With CB-Aloha, the time wasted on unsuccessfully transmitting RTS is equal to $\sigma_N$, i.e. $\Delta_{F,N}=\sigma_N$.
Fig. \ref{fig5} also gives the relation between time slot and data length in PB-Aloha and CB-Aloha:
\begin{equation}\label{data_L_relation}
	L_N=K\cdot L_P=M\cdot \sigma_N;~\sigma_P=L_P+\Delta_{S,P},
\end{equation}
where $K$ is the number of the data length transmitted by PB-Aloha at one transmission.
Due to the different slot length between $\sigma_N$ and $\sigma_P$, when given $\lambda_{N}$, it is normalized by the time slot $\sigma_N$ and subsequently rewritten in the unit of PB-Aloha time slot length $\sigma_P$, therefore, we have $\lambda_{P}=\lambda_{N} \frac{\sigma_P}{\sigma_N}$.

In the following, we will demonstrate the impact of $K$ and $L_P$ on the performance of CB-Aloha and PB-Aloha.
Combining \eqref{connection_Umaxp} and \eqref{packet_Umaxp}, when both networks are saturated and without life time constraint, we have $U_{\max}^{p,P} > U_{\max}^{p,N}$ if and only if the following inequality is satisfied:
\begin{equation}\label{compare_saturated}
	\begin{array}{ll}
		\frac{ \frac{KL_P(L_P+\Delta_{S,P})}{\sigma_N} -\sigma_N}{KL_P+\sigma_N(\delta-1)} < \frac{((n-1)P_W+P_T)p_{\text{m}}\ln p_\text{m}}{(P_T-P_W)\ln p_\text{m}-nP_W},
	\end{array}
\end{equation}
where $p_\text{m}$ is given by \eqref{connection_Umaxp}.
When both networks are unsaturated and without lifetime constraint, we have $U_{\max}^{p,P} > U_{\max}^{p,N}$ if and only if the following inequality is satisfied:
\begin{equation}\label{compare_unsaturated}
	\begin{array}{ll}
		\frac{K L_P p_L(L_P+\Delta_{S,P})}{\sigma_N^2+(KL_P+\sigma_N(\delta-1))p_L\sigma_N} < \exp\left( \mathbb{W}_0(-n\lambda_{P}) \right) ,
	\end{array}
\end{equation}
where $p_L$ is given by \eqref{T_gen}.

Compared to PB-Aloha, the transmission failures of CB-Aloha only involve RTS frames, at the cost of the overhead of connection establishment.
However, the benefits brought by the reduction of transmission failure may not always overweight the overhead, especially when there are few collisions in the network or the successful transmission time is small.
As Fig. \ref{Both_network_compare} illustrates, when both networks are unsaturated, nodes do not send data packets frequently, and there are few collisions in the networks. When $K$ is small, i.e., the successful transmission time of CB-Aloha is small, $U_{\max}^{p,P} < U_{\max}^{p,N}$ only when $L_P$ is large, where the packet transmission time of PB-Aloha is large and the energy wasting on failed transmission becomes high.
As $K$ grows, i.e., the successful transmission time of CB-Aloha increases, $U_{\max}^{p,P} < U_{\max}^{p,N}$ within a wider range of $L_P$ value.
When both networks are saturated, nodes frequently access the channel, resulting in more collisions. In this case, the region where $U_{\max}^{p,P} < U_{\max}^{p,N}$ has expanded compared to the unsaturated case.

\subsection{Case Study: 4-step RA-SDT versus 2-step RA-SDT }
To illustrate the practical application of the aforementioned analysis to real M2M networks, let's delve into the details of the 2-step RA-SDT and 4-step RA-SDT schemes, as outlined in the 3GPP Release 17\cite{3GPP_RASDT}.
Detailed schematic diagrams and parameters setting of these two schemes are introduced in \cite{ConnectionAloha}:	$\Delta_{S,P}=\Delta_{F,P}=6\text{ms}, \Delta_{S,N}=8\text{ms}, \sigma_N=2\text{ms}$. 
By setting $P_T=300\text{mW}, P_W=3\text{mW}, n=100$ and combining \eqref{data_L_relation}, \eqref{compare_saturated} and \eqref{compare_unsaturated}, we can give the threshold $K$, denoted as $K^*$ and $L_P$, denoted as $L_P^*$ which enable $U_{\max}^{p,P} > U_{\max}^{p,N}$ :

\begin{small}
    \begin{equation}\label{pratical network compare}
	\left\{\!\!
	\begin{array}{ll}
 		L_N < \frac{8.9356}{L_P+5.1774}
		\,\,\,\,\text{when both networks are saturated,}\\
		\frac{L_N p_L}{4+2p_L(L_N+6)} < \frac{\exp\left( \mathbb{W}_0(-\hat{\lambda}_N \frac{L_P+6}{2}) \right)}{L_P+6}\\ \,\,\,\,\,\,\,\,\,\,\,\,\,\,\,\,\,\,\,\,\,\,\,\,\,\,\,\,\,\,\,\,\,\,\,\,\,\,\,\,\,\,\,\,\,\,\text{when both networks are unsaturated}\\
	\end{array}\right.
\end{equation}
\end{small}

\noindent
where $p_L$ is given by \eqref{T_gen}.

When both networks are saturated and the data length of 2-step RA-SDT $L_P=0.5\text{ms}$, for example, according to \eqref{pratical network compare}, we have $U_{\max}^{p,P} > U_{\max}^{p,N}$ as long as $L_N< 1.57\text{ms} \approx 3L_P$.
It indicates that using the 4-step RA-SDT to transmit data of the same length as the 2-step RA-SDT will cause low energy efficiency. However, when $L_N > 1.57\text{ms}$, we have $U_{\max}^{p,N} > U_{\max}^{p,P}$, indicating that 4-step RA-SDT can improve energy efficiency by transmitting multiple packets during one connection which to some extent reduces the average overhead required to transmit each data packet. 

\section{Conclusion}
This letter considers the energy efficiency optimization of Aloha networks with a finite battery budget. We derived closed-form expression of lifetime and lifetime throughput of CB-Aloha and PB-Aloha. The analysis reveals that there is a tradeoff between lifetime and lifetime throughput. We maximize lifetime throughput under lifetime constraints by tuning the channel access probability in CB-Aloha and PB-Aloha. We derive the optimal operating regime of these two schemes. Then, we apply our analysis to practical RA-SDT networks.  Our analysis revealed that while the 4-step RA-SDT increases overhead during frequent small data transmissions, it still enhances energy efficiency by enabling the transmission of multiple data packets within one connection.

\appendices
\section{Proof of Lemma \ref{lemma1}}\label{app0}
The transmission states can be divided into successful transmission states and failed states. Note that $p_N$ denotes the probability of successful transmission. In each transmission attempt, with probability $p_N$, each node spends $M+\delta$ time slots in successful transmissions, and with probability $1-p_N$, it spends one time slot in collision. So the number of the successful transmission state and failed state satisfies the following relation:
\begin{small}
\begin{align}\label{Eq5}
	\frac{{n_{S}}}{n_{F}}=\frac{p_N}{1-p_N}. 
\end{align}
\end{small}

\noindent
Since the CB-Aloha network can be analyzed by a request-queue model in \cite{ConnectionAloha}, the mean request service rate of each node, $\mu_{r}$, which is defined as a ratio of the number of successful requests and each node's lifetime, can be written as:
\begin{small}
\begin{align}\label{Eq6}
	\mu_{r}=\frac{n_{S}}{n_{W}+n_{F}+n_{S}(M+\delta)}.
\end{align}
\end{small}

\noindent
The offered load $\rho$ of each node's data queue is then given by
\begin{small}
    \begin{align}\label{Eq7}
        \rho=\frac{\lambda_{r}}{\mu_{r}}=\frac{\lambda_N\left(n_{W}+n_{F}+n_{S}(M+\delta)\right)}{Mn_{S}},
\end{align}
\end{small}

\noindent
where $\lambda_N$ is the packet arrival rate of each node. When the network is unsaturated with $\rho<1$, the offered load equals the probability that the nodes' queue is not empty. We then have
\begin{small}
    \begin{align}\label{Eq8}
\frac{{n_{W}+n_{F}+n_{S}(M+\delta)}}{{n_{I}}}=\frac{\rho}{1-\rho}.
\end{align}
\end{small}

\noindent
By combining \eqref{Eq2}, \eqref{Eq3}, \eqref{Eq5}, \eqref{Eq7} and \eqref{Eq8}, the expected lifetime of each node $T_N$ in unsaturated situation can be obtained.
When the network becomes saturated with $\rho\geq1$, we have $n_{I}=0$. In this case, the mean service rate of each node's queue $\mu_{r}$ equals its throughput, then


\begin{small}
    \begin{align}\label{Add2}
\mu_{r}=\frac{p_N\ln{p_N}}{np_N\ln{p_N}\left( M+\delta-1\right) -n }.
\end{align}
\end{small}

\noindent
By combining \eqref{Eq2}, \eqref{Eq3}, \eqref{Eq5}, \eqref{Eq6} and \eqref{Add2}, then $T_N$ in saturated situation can be obtained.




\section{Proof of Theorem \ref{t1}}\label{app1}
To get the optimal lifetime throughput with the lifetime constraint. We first propose a Lemma.

\begin{lemma}\label{lemmaApp1}
The maximum lifetime throughput without any constraints $U_{\max}^{p,N,T_0=0}$ is given by
\begin{equation}\label{M_maxp0}
U^{p,N,T_0=0}_{\max}=\left\{\!\!\!
\begin{array}{ll}
\frac{E/\sigma_N}{\frac{ 1+(M+\delta-1)p_L }{M p_L}(P_{T}-P_{W})+\frac{P_{W}}{\lambda_N}}
\,\,\,\,\,\,\,\,\,\,\,\,\,\,\text{if}\,\,p_\text{m}\le p_L\\
\frac{E/\sigma_N}{\frac{((n-1)P_{W}+P_{T})(M+\delta-1)}{M}+\frac{P_{T}-P_{W}}{Mp_\text{m}}-\frac{nP_{W}}{Mp_\text{m}\ln{p_\text{m}}}}\\
\,\,\,\,\,\,\,\,\,\,\,\,\,\,\,\,\,\,\,\,\,\,\,\,\,\,\,\,\,\,\,\,\,\,\,\,\,\,\,\,\,\,\,\,\,\,\,\,\,\,\,\,\,\,\,\,\,\,\,\,\,\,\,\,\,\,\,\,\,\,\,\,\,\,\,\,\,\,\,\text{otherwise,}
\end{array}\right.
\end{equation}
where $p_\text{m}=\exp\left\{  \frac{n-\sqrt{n^2+4n\left( \frac{P_{T}}{P_{W}}-1 \right)}}{2\left( \frac{P_{T}}{P_{W}}-1 \right) } \right\}$.
\end{lemma}

\begin{IEEEproof}
Let
\begin{small}
\begin{equation}
    f(p_N)=\frac{E/\sigma_N}{\frac{((n-1)P_{W}+P_{T})(M+\delta-1)}{M}+\frac{P_{T}-P_{W}}{Mp_N}-\frac{nP_{W}}{Mp_N\ln{p_N}}}.
\end{equation}
\end{small}

\noindent
If $p_N\notin S(p_N,\hat{\lambda}_N)$, 
then we have $U_N=f(p_N)$. It can be proved that $f(p_N)$ monotonically increases as $p_N$ increases if $p_N < p_\text{m}$, and monotonically decreases as $p_N$ increases if $p_N > p_\text{m}$. $f(p_N)$ is then maximized when $p_N=p_\text{m}$. It is clear that if $\frac{\hat{\lambda}_N}{M-\hat{\lambda}_N (M+\delta-1)} \geq e^{-1}$, then the network will become saturated and $U_N$ is maximized when $p_N=p_\text{m}$. In the following, we focus on the condition of $ 0<\frac{\hat{\lambda}}{M-\hat{\lambda}_N (M+\delta-1)} < e^{-1}$. Notice that as $P_T\geq P_W$, we have $p_\text{m}\ge \exp\{-1\} \ge\exp{\left( {\it {\mathbb W}}_{-1}(-\frac{\hat{\lambda}_N}{M-\hat{\lambda}_N(M+\delta-1)})\right) } $. We then divide the discussion into two cases:

\vspace{0.6em}

\noindent \textbf{Case 1:}
$p_\text{m}\le \exp{\left( {\it {\mathbb W}}_{0}(-\frac{\hat{\lambda}_N}{M-\hat{\lambda}_N (M+\delta-1)})\right) } = p_L$: \\
This condition is equivalent to $\lambda_N \le \lambda_M^N$ according to \eqref{connection_Umaxp}. 
Due to the monotonicity of $f(p_N)$, we have $f(p_N)$ monotonically increases as $p_N$ increases when $p_N < \exp{\left( {\it {\mathbb W}}_{-1}(-\frac{\hat{\lambda}_N}{M-\hat{\lambda}_N (M+\delta-1)})\right) }$, and decreases as $p_N$ increases when $p_N > \exp{\left( {\it {\mathbb W}}_{0}(-\frac{\hat{\lambda}_N}{M-\hat{\lambda}_N (M+\delta-1)})\right) }$. \\
Note that: 
\begin{small}
\begin{equation}
    \frac{E/\sigma_N}{\frac{ 1+(M+\delta-1)p_N }{Mp_N}(P_{T}-P_{W})+\frac{P_{W}}{\lambda_N}}=f(p_N),
\end{equation}
\end{small}

\noindent when 
\begin{small}
    \begin{equation}
        p_N=\exp{\left( {\it {\mathbb W}}_{0}(-\frac{\hat{\lambda}_N}{M-\hat{\lambda}_N (M+\delta-1)})\right) } 
    \end{equation}
\end{small}

\noindent or
\begin{small}
    \begin{equation}
        p_N=\exp{\left( {\it {\mathbb W}}_{-1}(-\frac{\hat{\lambda}_N}{M-\hat{\lambda}_N(M+\delta-1)})\right) },
    \end{equation}
\end{small}

\noindent and $\frac{E/\sigma_N}{\frac{ 1+(M+\delta-1)p_N }{Mp_N}(P_{T}-P_{W})+\frac{P_{W}}{\lambda_N}}$ is a monotonically non-decreasing function of $p_N$. Therefore, $U_N$ is maximized when $p_N=\exp{\left( {\it {\mathbb W}}_{0}(-\frac{\hat{\lambda}_N}{M-\hat{\lambda}_N(M+\delta-1)})\right) }$.

\vspace{0.6em}

\noindent \textbf{Case 2:} $p_\text{m}>\exp{\left( {\it {\mathbb W }}_{0}(-\frac{\hat{\lambda}_N}{M-\hat{\lambda}_N(M+\delta-1)})\right) } =p_L$:\\
This condition is equivalent to $\lambda_N > \lambda_M^N$ according to \eqref{connection_Umaxp}. 
In this case, we have $U_N=f(p_N)$ if $p_N\notin S(p_N,\hat{\lambda}_N)$, which is maximized when $p_N=p_\text{m}$. If $p_N\in S(p_N,\hat{\lambda}_N)$, then we have $U_N<f(p_\text{m})$.
So $U_N$ is maximized when  $p_N=p_\text{m}$.
\end{IEEEproof}
\vspace{0.7em}

\textbf{Now let us take into consideration the lifetime constraint} of each CB-Aloha node $T_N\ge T_0^N$. According to \eqref{T_gen}, we have: $T_{\max}^N= \max_{p_N}T_N=\frac{E/\sigma_N}{P_W}$. If the lifetime constraint $T_0^N>\frac{E/\sigma_N}{P_W}$, then the optimization problem \eqref{Mc_did1} is not feasible.
If $P_T=P_W$, then we have $T_N=\frac{E/\sigma_N}{P_W}$ according to \eqref{T_gen}, in this case, if the lifetime constraint $T_0^N\le\frac{E/\sigma_N}{P_W}$, then the optimization problem \eqref{Mc_did1} becomes unconstrained optimization, and the solution is given by \eqref{M_maxp0}. When $P_T>P_W$ and $T_0^N\le\frac{E/\sigma_N}{P_W}$, we divide the discussion into two cases:
\vspace{0.6em}

\noindent \textbf{Case 1:} $p_\text{m}\le \exp{\left( {\it {\mathbb W }}_{0}(-\frac{\hat{\lambda}_N}{M-\hat{\lambda}_N(M+\delta-1)})\right) }=p_L$: \\
In this case, we have:
\begin{small}
    \begin{equation}
        T_N\left( p_L\right) =\frac{E/\sigma_N}{\frac{\left( 1+(M+\delta-1)p_L\right) \lambda_N}{Mp_L}(P_{T}-P_{W})+P_{W}},
    \end{equation}
\end{small}
 
\noindent according to Lemma \ref{lemma1}, where $T_N(\cdot)$ is a function of $p_N$ given by \eqref{T_gen}.
If $T_0^N \le T_N\left( p_L\right)$, then
\begin{small}
    \begin{equation}
        p_N=\exp{\left( {\it {\mathbb W }}_{0}(-\frac{\hat{\lambda}_N}{M-\hat{\lambda}_N(M+\delta-1)})\right) }\in\{p_N|T_0^N\le T_N\},
    \end{equation}
\end{small}

\noindent
indicating that $p_N=\exp{\left( {\it {\mathbb W }}_{0}(-\frac{\hat{\lambda}_N}{M-\hat{\lambda}_N(M+\delta-1)})\right) }$ lies in the feasible region of the optimization problem \eqref{Mc_did1}. According to Case 1 of the unconstrained optimization problem, we have:
\begin{small}
\begin{equation}
    U^{p,N}_{\max}=\frac{E/\sigma_N}{\frac{ 1+(M+\delta-1)\exp{\left( {\it {\mathbb W }}_{0}(-\frac{\hat{\lambda}_N}{M-\hat{\lambda}_N(M+\delta-1)})\right) } }{M\exp{\left( {\it {\mathbb W }}_{0}(-\frac{\hat{\lambda}_N}{M-\hat{\lambda}_N(M+\delta-1)})\right) }}(P_{T}-P_{W})+\frac{P_{W}}{\lambda_N}}
\end{equation}
\end{small}

\noindent
which is achieved when $p_N=\exp{\left( {\it {\mathbb W }}_{0}(-\frac{\hat{\lambda}_N}{M-\hat{\lambda}_N(M+\delta-1)})\right) }$.
If $T_0^N>T_N\left( \exp{\left( {\it {\mathbb W }}_{0}(-\frac{\hat{\lambda}_N}{M-\hat{\lambda}_N(M+\delta-1)})\right) }\right)=  \\$
\begin{small}
    \begin{equation}
    \frac{E/\sigma_N}{\frac{\left( 1+(M+\delta-1)\exp{\left( {\it {\mathbb W }}_{0}(-\frac{\hat{\lambda}_N}{M-\hat{\lambda}_N(M+\delta-1)})\right) }\right) \lambda_N}{M\exp{\left( {\it {\mathbb W }}_{0}(-\frac{\hat{\lambda}_N}{M-\hat{\lambda}_N(M+\delta-1)})\right) }}(P_{T}-P_{W})+P_{W}}, 
\end{equation}
\end{small}

\noindent
then $T_N\ge T_0^N$ is equivalent to $p_N\ge p_c^N>\exp{\left( {\it {\mathbb W }}_{0}(-\frac{\hat{\lambda}_N}{M-\hat{\lambda}_N(M+\delta-1)})\right) }$, as $T_N$ monotonically increases as $p_N$ increases. As $p_\text{m}\le\exp{\left( {\it {\mathbb W}}_{0}(-\frac{\hat{\lambda}}{M-\hat{\lambda}(M+\delta-1)})\right) }$, $U_N$ monotonically decreases as $p_N$ increases when $p_N>\exp{\left( {\it {\mathbb W }}_{0}(-\frac{\hat{\lambda}_N}{M-\hat{\lambda}_N(M+\delta-1)})\right) }$. As a result, we have:
\begin{equation}
    U^{p,N}_{\max}=\frac{E/\sigma_N}{\frac{((n-1)P_{W}+P_{T})(M+\delta-1)}{M}+\frac{P_{T}-P_{W}}{Mp_{c}}-\frac{nP_{W}}{Mp_{c}\ln{p_{c}}}},
\end{equation}
which is achieved when $p_N=p_c^N$.  $p_{c}^N$ is the solution of the equation \eqref{solvePc}.
\vspace{0.6em}

\noindent \textbf{Case 2:} $p_\text{m}>\exp{\left( {\it {\mathbb W }}_{0}(-\frac{\hat{\lambda}_N}{M-\hat{\lambda}_N(M+\delta-1)})\right) }=p_L$:\\
In this case, we have:
\begin{small}
    \begin{equation}
    T_N({p_\text{m}})=\frac{\frac{E}{\sigma_N}(1-p_\text{m}\ln{p_\text{m}}(M+\delta-1))}{P_{W}-\frac{((n-1)P_{W}+P_{T})(M+\delta-1)p_\text{m}\ln{p_\text{m}}}{n}-\frac{(P_{T}-P_{W})\ln{p_\text{m}}}{n}},
\end{equation}
\end{small}

\noindent
according to Lemma \ref{lemma1}, if $T_0^N\le T_N({p_\text{m}})$, then 
\begin{equation}
    p_N=p_\text{m}\in\{p_N|T_0^N\le T_N\},
\end{equation}

\noindent indicating that $p_N=p_\text{m}$ lies in the feasible region of the optimization problem \eqref{Mc_did1}. According to Case 2 of the unconstrained optimization problem, we have:

\begin{equation}
    U^{p,N}_{\max}=\frac{E/\sigma_N}{\frac{((n-1)P_{W}+P_{T})(M+\delta-1)}{M}+\frac{P_{T}-P_{W}}{Mp_\text{m}}-\frac{nP_{W}}{Mp_\text{m}\ln{p_\text{m}}}}, 
\end{equation}
which is achieved when $p_N=p_\text{m}$.
If $T_0^N> T_N({p_\text{m}})$, then $T_N\ge T_0^N$ is equivalent to $p_N\ge p_c^N> p_\text{m}$, as $T_N$ monotonically increases as $p_N$ increases. As $p_\text{m} > \exp{\left( {\it {\mathbb W}}_{0}(-\frac{\hat{\lambda}_N}{M-\hat{\lambda}_N(M+\delta-1)})\right) }$, $U_N$ monotonically decreases as $p_N$ increases when $p_N>p_\text{m}$. As a result,  we have:
\begin{equation}
    U^{p,N}_{\max}=\frac{E/\sigma_N}{\frac{((n-1)P_{W}+P_{T})(M+\delta-1)}{M}+\frac{P_{T}-P_{W}}{Mp_{c}}-\frac{nP_{W}}{Mp_{c}\ln{p_{c}}}},
\end{equation}
which is achieved when $p_N=p_c^N$.  $p_{c}^N$ is the solution of the equation \eqref{solvePc}.
Then by using the explicit relation between $p_N$ and $q$, we can easily obtain the optimal transmission probability $q_{\max}^N$, given by \eqref{qM_N}.

\bibliographystyle{IEEEtran}
\bibliography{IEEEabrv,Fn_Ref}

\begin{thebibliography}{10}
\providecommand{\url}[1]{#1}
\csname url@samestyle\endcsname
\providecommand{\newblock}{\relax}
\providecommand{\bibinfo}[2]{#2}
\providecommand{\BIBentrySTDinterwordspacing}{\spaceskip=0pt\relax}
\providecommand{\BIBentryALTinterwordstretchfactor}{4}
\providecommand{\BIBentryALTinterwordspacing}{\spaceskip=\fontdimen2\font plus
\BIBentryALTinterwordstretchfactor\fontdimen3\font minus
  \fontdimen4\font\relax}
\providecommand{\BIBforeignlanguage}[2]{{%
\expandafter\ifx\csname l@#1\endcsname\relax
\typeout{** WARNING: IEEEtran.bst: No hyphenation pattern has been}%
\typeout{** loaded for the language `#1'. Using the pattern for}%
\typeout{** the default language instead.}%
\else
\language=\csname l@#1\endcsname
\fi
#2}}
\providecommand{\BIBdecl}{\relax}
\BIBdecl

\bibitem{EEsurvey2022}
D.~López, A.~De~Domenico, N.~Piovesan, G.~Xinli, H.~Bao, S.~Qitao, and
  M.~Debbah, ``{A Survey on 5G Radio Access Network Energy Efficiency: Massive
  MIMO, Lean Carrier Design, Sleep Modes, and Machine Learning},'' \emph{{IEEE}
  Commun. Surveys Tuts.}, vol.~24, no.~1, pp. 653--697, 1st Quart., 2022.

\bibitem{EEsurvey2016}
R.~Mahapatra, Y.~Nijsure, G.~Kaddoum, N.~Ul~Hassan, and C.~Yuen, ``{Energy
  Efficiency Tradeoff Mechanism Towards Wireless Green Communication: A
  Survey},'' \emph{{IEEE} Commun. Surveys Tuts.}, vol.~18, no.~1, pp. 686--705,
  1st Quart., 2016.

\bibitem{3GPP_RASDT}
\emph{{5G; NR; Medium Access Control (MAC) protocol specification (Release
  17)}}, document TS 38.321 V17.0.0, 3GPP, May 2022.

\bibitem{noma1}
H.~Kong, M.~Lin, L.~Han, W.-P. Zhu, Z.~Ding, and M.-S. Alouini, ``{Uplink
  Multiple Access With Semi-Grant-Free Transmission in Integrated
  Satellite-Aerial-Terrestrial Networks},'' \emph{{IEEE} J. Sel. Areas
  Commun.}, vol.~41, no.~6, pp. 1723--1736, June 2023.

\bibitem{noma2}
Y.~Guo, M.~Lin, H.~Kong, M.~Cheng, and W.-P. Zhu, ``{NOMA Assisted
  Semi-Grant-Free Transmission Scheme in Satellite Systems},'' \emph{{IEEE}
  Commun. Lett.}, vol.~27, no.~8, pp. 2122--2126, Aug. 2023.

\bibitem{gao2023random}
Y.~Gao, W.~Zhan, and L.~Dai, ``{Random Access: Connection-Free or
  Connection-Based?}'' in \emph{From 5g To 6g and Beyond: The 7 Cs Of Future
  Communications}.\hskip 1em plus 0.5em minus 0.4em\relax World Scientific,
  2023, pp. 105--140.

\bibitem{Dai_22TON}
L.~Dai, ``{A Theoretical Framework for Random Access: Stability Regions and
  Transmission Control},'' \emph{IEEE/ACM Trans. Networking}, vol.~30, no.~5,
  pp. 2173--2200, Oct. 2022.

\bibitem{19GaoDai}
Y.~Gao and L.~Dai, ``{Random Access: Packet-Based or Connection-Based?}''
  \emph{IEEE Trans. Wireless Commun.}, vol.~18, no.~5, pp. 2664--2678, Mar.
  2019.

\bibitem{ConnectionAloha}
X.~Zhao and L.~Dai, ``{Connection-Based Aloha: Modeling, Optimization, and
  Effects of Connection Establishment},'' \emph{IEEE Trans. Wireless Commun.},
  vol.~23, no.~2, pp. 1008--1023, Feb. 2024.

\bibitem{IRSA_EE}
Z.~Chen, Y.~Feng, Z.~Tian, Y.~Jia, M.~Wang, and T.~Q.~S. Quek, ``{Energy
  Efficiency Optimization for Irregular Repetition Slotted ALOHA-Based Massive
  Access},'' \emph{IEEE Wireless Commun. Lett.}, vol.~11, no.~5, pp. 982--986,
  May 2022.

\bibitem{Aloha}
L.~Dai, ``{Stability and Delay Analysis of Buffered Aloha Networks},''
  \emph{IEEE Trans. Wireless Commun.}, vol.~11, no.~8, pp. 2707--2719, Aug.
  2012.

\bibitem{zhq_EE}
X.~Sun, H.~Zhang, W.~Zhan, X.~Wang, and X.~Chen, ``{How to Survive 10 Years’
  Life Time for Machine Type Devices: A Study of Random Access With
  Sleeping-Awake Cycle},'' \emph{IEEE Trans. Commun.}, vol.~71, no.~11, pp.
  6727--6744, Nov. 2023.

\bibitem{CBAloha_TCOMM}
H.~Zhou, Y.~Deng, L.~Feltrin, and A.~Höglund, ``{Analyzing Novel Grant-Based
  and Grant-Free Access Schemes for Small Data Transmission},'' \emph{IEEE
  Trans. Commun.}, vol.~70, no.~4, pp. 2805--2819, Feb. 2022.

\end{thebibliography}
\end{document}